\documentclass[journal]{vgtc}                % final (journal style)
\ifpdf%                                % if we use pdflatex
  \pdfoutput=1\relax                   % create PDFs from pdfLaTeX
  \usepackage{graphicx}                % allow us to embed graphics files
  \DeclareGraphicsExtensions{.pdf,.png,.jpg,.jpeg} % for pdflatex we expect .pdf, .png, or .jpg files
\else%                                 % else we use pure latex
  \usepackage{graphicx}                % allow us to embed graphics files
  \DeclareGraphicsExtensions{.eps}     % for pure latex we expect eps files
\fi%

\graphicspath{{figures/}{pictures/}{images/}{./}} % where to search for the images

\usepackage{microtype}                 % use micro-typography (slightly more compact, better to read)
\PassOptionsToPackage{warn}{textcomp}  % to address font issues with \textrightarrow
\usepackage{textcomp}                  % use better special symbols
\usepackage{mathptmx}                  % use matching math font
\usepackage{times}                     % we use Times as the main font
\usepackage{cite}                      % needed to automatically sort the references
\usepackage{tabu}                      % only used for the table example
\usepackage{booktabs}                  % only used for the table example
\usepackage{paralist}
\usepackage{enumitem}
\usepackage{multirow}
\usepackage{color}
\usepackage{amsmath}
\usepackage{bm}
\usepackage{xspace}
\usepackage{soul}
\usepackage{setspace}

\newcommand{\name}{Causality Explorer\xspace}
\newcommand{\q}[1]{\textit{``#1''}}

\newcommand{\minisection}[1]{\textbf{#1:}}

\newcommand\independent{\protect\mathpalette{\protect\independenT}{\perp}}
\def\independenT#1#2{\mathrel{\rlap{$#1#2$}\mkern2mu{#1#2}}}

\onlineid{1041}

\vgtccategory{Research}
\vgtcpapertype{Application \& Design Study}

\title{A Visual Analytics Approach for Exploratory Causal Analysis: Exploration, Validation, and Applications}

\author{Xiao Xie, Fan Du, and Yingcai Wu}
\authorfooter{
\item
 X. Xie and Y. Wu are with the State Key Lab of CAD\&CG, Zhejiang University. E-mail: \{xxie, ycwu\}@zju.edu.cn.
\item
 F. Du is with Adobe Research and is the corresponding author. E-mail: fdu@adobe.com.
}

\shortauthortitle{Biv \MakeLowercase{\textit{et al.}}: Global Illumination for Fun and Profit}
\abstract{Using causal relations to guide decision making has become an essential analytical task across various domains, from marketing and medicine to education and social science. While powerful statistical models have been developed for inferring causal relations from data, domain practitioners still lack effective visual interface for interpreting the causal relations and applying them in their decision-making process. Through interview studies with domain experts, we characterize their current decision-making workflows, challenges, and needs. Through an iterative design process, we developed a visualization tool that allows analysts to explore, validate, and apply causal relations in real-world decision-making scenarios. The tool provides an uncertainty-aware causal graph visualization for presenting a large set of causal relations inferred from high-dimensional data. On top of the causal graph, it supports a set of intuitive user controls for performing what-if analyses and making action plans. We report on two case studies in marketing and student advising to demonstrate that users can effectively explore causal relations and design action plans for reaching their goals.

} % end of abstract


\CCScatlist{ % not used in journal version
 \CCScat{K.6.1}{Management of Computing and Information Systems}%
{Project and People Management}{Life Cycle};
 \CCScat{K.7.m}{The Computing Profession}{Miscellaneous}{Ethics}
}

\teaser{
  \centering
  \includegraphics[width=1\linewidth]{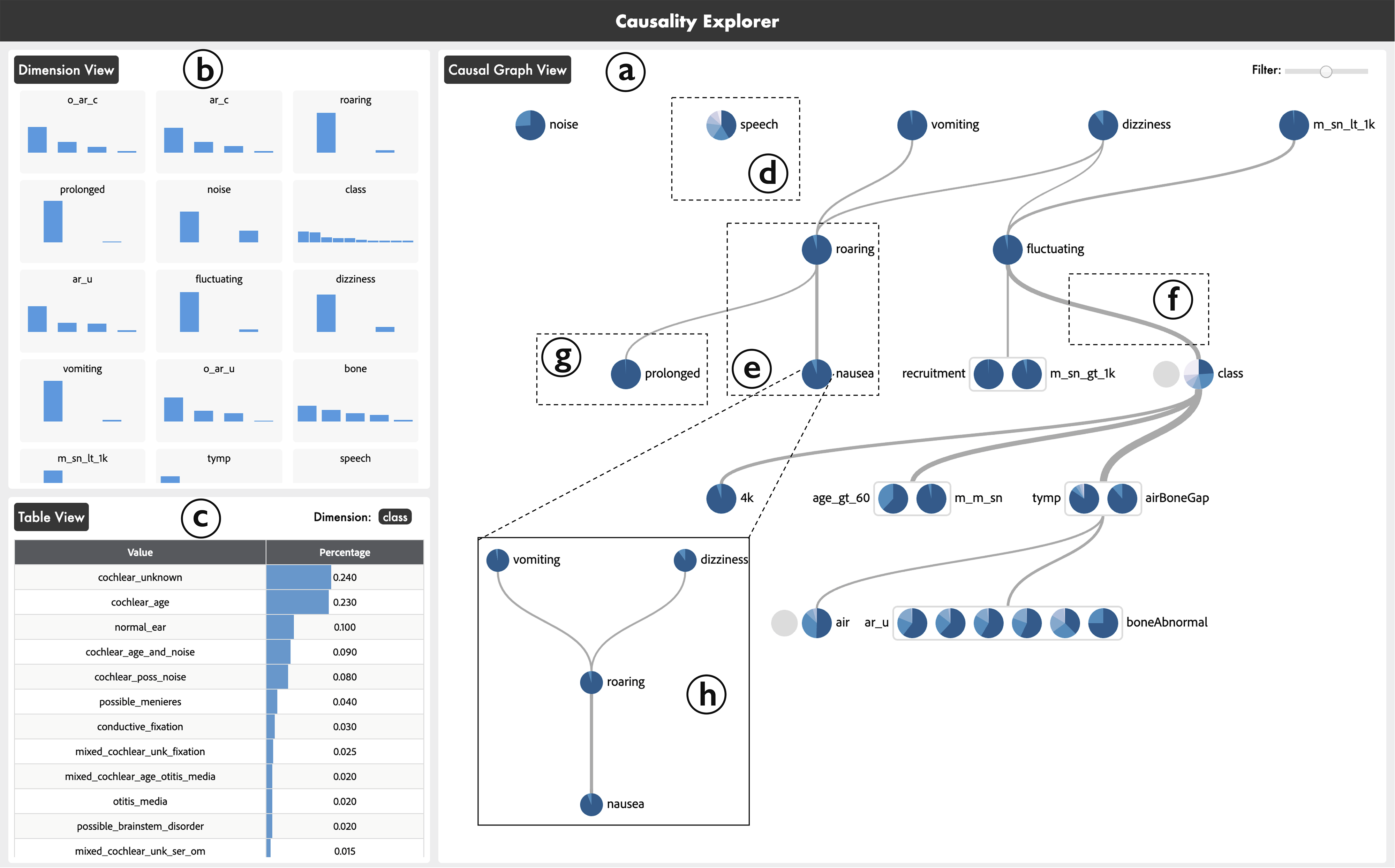}
  \caption{The user interface of Causality Explorer demonstrated with a real-world audiology dataset that consists of 200 rows and 24 dimensions~\cite{audiology}. (a) A scalable causal graph layout that can handle high-dimensional data. (b) Histograms of all dimensions for comparative analyses of the distributions. (c) Clicking on a histogram will display the detailed data in the table view. (b) and (c) are coordinated to support what-if analyses. In the causal graph, each node is represented by a pie chart (d) and the causal direction (e) is from the upper node (cause) to the lower node (result). The thickness of a link encodes the uncertainty (f). Nodes without descendants are placed on the left side of each layer to improve readability (g). Users can double-click on a node to show its causality subgraph (h).}
	\label{system}
}

\vgtcinsertpkg

\begin{document}

%% The ``\maketitle'' command must be the first command after the
%% ``\begin{document}'' command. It prepares and prints the title block.

%% the only exception to this rule is the \firstsection command
% \firstsection{Introduction}

\maketitle

\begin{spacing}{0.96}

\section{Introduction} \label{sec:introduction}

Causal reasoning is a common task in data analysis and decision making. Doctors may want to identify the root cause of a disease symptom while marketers would hope to understand which customer segments contributed the most to their revenue. Due to the high cost of controlled experiments, most of the existing analytics systems apply correlation analysis to derive such causal conclusions. However, the fact that correlation is not causation motivates the involvement of causal analysis, which aims to infer causal relations from observational data. 

Two categories of exploratory causal analysis models, namely, the constraint-based ones (e.g., SGS\cite{sgs/spirtes1991}, PC\cite{pc/spirtes1991}) and the score-based ones (GES\cite{jmlr/Chickering02a}, F-GES\cite{ijdsa/RamseyGSG17}), have been experimented for causal discovery. These methods apply different detection approaches but share the same output, i.e., a causal graph where the nodes encode the data dimensions and edges encode the causal directions. Numerous high-value applications can be developed on top of these causal graphs. For example, in digital marketing, analysts can use the causal graph to identify important factors leading to purchase orders or simulate the outcomes of different campaign strategies.

In recent years, researchers have designed tailored interactive visualization systems for exploratory causal analysis. However, two main challenges remain to be resolved to fully utilize the detected causal graph for real-world applications. First, when detecting causal relations in a high-dimensional dataset, the state-of-the-art solution is F-GES model, which applies a greedy search for the causal relations. Although the detection process can be highly accelerated, this raises an uncertainty issue, i.e., the model cannot ensure the quality of the detected causal relation. How to estimate and present the uncertainty of the detected causal relations remains to be resolved.
The second challenge is the lack of interactive tools for utilizing the causal graph. Wang et al.~\cite{tvcg/WangM16} have developed a visualization system for presenting the causal graph and proposed interactions that can support the diagnosis of the detected causal graph. Despite the usefulness, the system was not designed to handle a large causal graph, which is commonly seen in domain datasets such as marketing, healthcare, and education. Moreover, rather than exploring the causal graph, how to best integrate human knowledge with the causal graph for decision-making applications like simulations and attributions remains an open research question.

Consider a campaign use case scenario. A marketer Bob is designing a campaign for promoting the subscription renewal of a group of customers. Given the constraints of budgets, Bob hopes to only use a few efficient marketing channels for the campaign. By applying correlation analysis on historical marketing data, Bob identifies a set of channels that are highly correlated with the past renewals of the group. However, the selection is still difficult, since correlations do not imply the pure effect of each channel. For example, among customers who received sales emails and renewed, it is misleading to say all of the renewals attribute to the emails, because many of the customers may have already established a purchase intent before from other channels such as social media. Besides, Bob also struggles with how to convince the stakeholders of his campaign plan, since the performance of a plan is hard to estimate without running expensive A/B testings and a large number of testings will be needed given that Bob has no clue how to narrow down the channel selection.

In this paper, we seek to address these gaps in the applications of causal analysis by designing an interactive visualization system with domain practitioners. We first interview real-world data analysts to understand their fundamental design needs for applying causal analysis. Next, we adopt the state-of-the-art causal discovery model to handle the scalability issue raised by data dimensions and also extract the uncertainty of the detected causality. Finally, we design a scalable causal graph visualization to enable analysts to explore the causal relations of high-dimensional data. Facet views and interactions are tailored to support analysts conducting what-if analysis on the causal graph. We evaluate the system on datasets from two different domains and report on two case studies with practitioners from education and digital marketing.
The direct contributions of this work are:
\begin{compactitem}
    \item A set of 7 design needs collected through interviews with 5 domain experts for visualizing large causal graphs and conducting what-if analysis.
    \item The design and implementation of an interactive visual analytics system, \name, for achieving practical causal analysis by supporting (1) uncertainty aware visualization of large-scale causal relations and (2) interactive what-if analysis and action plan simulation.
    \item An evaluation through case studies with domain practitioners to analyze education and digital marketing datasets.
\end{compactitem}

\section{Related Work} \label{sec:related_work}
In this section, we survey and discuss related literature around the discovery, visualizations, and applications of causal relations.

\subsection{Algorithms for Discovering Causal Relations}

The goal of causal discovery \cite{causal1, causal2} is to infer causal relations from a multi-dimensional dataset. Causal relations are commonly modeled as a Directed Acyclic Graph (DAG), where a node represents a data dimension and a link represents the dependency between two connected dimensions~\cite{ijon/Shanmugam01}. The arrows of the links indicate the direction of the cause-effect relationship. Existing causal discovery algorithms can be roughly grouped into two categories: constraint-based and score-based. Constraint-based algorithms, including SGS~\cite{sgs/spirtes1991} and PC~\cite{pc/spirtes1991, colombo2014order}, start with a fully connected graph and eliminate the links by performing conditional independence (CI) tests for each pair of dimensions. This process requires exponential numbers of CI tests, which is not scalable for large industry-level dataset. To scale up, GES~\cite{jmlr/Chickering02a}, a representative of the score-based algorithm, proposes a scoring function to estimate a DAG's fit to the dataset and transforms the detection problem to a greedy search problem. Ramsey et.al.~\cite{ijdsa/RamseyGSG17} further accelerated this method and proposed F-GES. By introducing additional assumptions and parallel computation techniques, F-GES can handle the causal discovery of high-dimensional data. In this paper, we apply F-GES for detecting the causal relations.

\subsection{Visualizations of Causal Relations}

Effectively presenting the causal graphs is critical for helping analysts interpret the causal relations. Based on a literature review, we summarize existing works into two categories: studies of causality perception in visualizations and visual analytics systems for exploratory causal analysis. In the causality perception category, Kadaba et.al.~\cite{tvcg/KadabaIL07} conducted experiments to evaluate the efficiency of static and animated graph visualizations on encoding causal information, such as the strength, the direction, and the causal effect (positive or negative). Bae et.al.~\cite{cgf/BaeHR17} examined whether a sequential graph layout can help users more easily realize the indirect causality and identify the root cause. Rather than showing the causality detected from statistical models, Yen et.al.~\cite{cgf/YenPF19} used bar charts to visualize the data and studied the performance (e.g., accuracy) of making causal inference with visualizations. Xiong et.al.~\cite{tvcg/XiongSHF20} studied the level of causality revealed by visualizations and found that users tend to draw causal conclusions rather than correlations when data is presented by high aggregated visualizations (e.g., bar charts). These empirical studies of causal visualizations provide useful design guidelines for our visual analytics system.

In real scenarios, it is often difficult to directly apply causal models to address domain problems without interactive tools. Different visual analytics systems are therefore proposed to integrate human intelligence into the causal analysis. Elmqvist and Tsigas~\cite{infovis/ElmqvistT03} presented a technique called Animated Growing Polygons for visualizing the causal relations between event sequences. Wang and Muller~\cite{tvcg/WangM16} introduced a system that integrated automatic causal discovery algorithms and visualizations. Users can inspect the detected causal graph and validate the causal links with interactions and statistical evidence. They further addressed the data subdivision problem in causal analysis with visualizations~\cite{ieeevast/WangM17}, i.e., users can create causal graphs for different subgroups of data and obtain insights by identifying the different causal relations among the subgroups. Although existing works have extensively investigated how to support the exploration of a causal graph, the graphs being evaluated are usually much smaller than those in real-world applications.

As a trade-off between speed and accuracy, score-based causal discovery algorithms (e.g., F-GES) are commonly applied by domain practitioners, which extracts an approximated large causal graph where each causal link is associated with a model uncertainty. How to visually present the uncertainty of a causal graph is therefore important for deriving trustworthy insights. Visualizing and communicating uncertainty \cite{tvcg/HullmanQCKK19} in graphs \cite{graph_uncertainty1, tvcg/GuoHL15, tvcg/SchulzNGDBW17, ShararaSNGS11} have received great attention in recent years. Wang et.al.~\cite{tvcg/WangSAZZYQ16} analyzed the uncertainty issues raised by graph layouts. Schulz et.al.~\cite{tvcg/SchulzNGDBW17} proposed a force-directed based visualizations to present a probabilistic graph model. Among these various works, Guo et.al.~\cite{tvcg/GuoHL15} studied the visualization of uncertainty within edges, which is most related to our work. They have evaluated the effectiveness of different visual encodings on presenting edge uncertainties with common graph tasks. However, their evaluations focused on the visualization of un-directed graphs while in causal analysis, each causal graph is assumed to be a DAG and the directions of the edges are important for interpreting the results. In this work, we address this gap by exploring the design space of applying uncertainty visualization techniques to directed causal graphs.

\subsection{Applications of Causal Relations}

% \task{FAN: change title. may be this title is too general.}
% \task{XIAO: existing techniques for attribution and what-if analysis. Causal techniques are better for these tasks because it is more accurate and more explainable, etc. But it has not been sufficiently studied.}

Researchers of various domains, such as digital marketing\cite{kdd/AbeVAS04, marketing1}, sports\cite{sportswhatif/eswa/VracarSK16, sportswhatif/tkdd/MeloALF12, sportswhatif/gabel2012random, sportswhatif/kdd/ChenJ16, sportswhatif/epjds/MerrittC14, sportswhatif/icdm/PeelC15}, and healthcare\cite{wsc/BrailsfordCJ17, health1}, have proposed statistical models to perform what-if analysis on data. Visual techniques and interactive tools \cite{homefinder, smartadp, airvis} have been developed to provide user-friendly interfaces for these models. A useful scenario in what-if is changing the feature value of a prediction model and inspect the updated model results for model comprehension\cite{iforest,modeltracker,gamut,liu2017towards}. Similarly, Prospector\cite{chi/KrausePN16} allows users to change the feature values of an instance and explore how this change affects the probability of classifications.
% (difference: individual and group, numerical and categorical, application scenarios: low-level and high-level)
Recently, focusing on the fairness issue, Wexler et.al. proposed WIT\cite{WIT} for conducting what-if analysis with machine learning models. With the aid of tailored interactions, users can test the machine learning models with different inputs and therefore obtain a better understanding of the model performance and the mechanism. 

In addition to the model comprehension, researchers have also studied how to apply what-if for addressing domain-specific problems\cite{cgf/LuGHGM17}. For example, in the domain of sports, what-if analyses are usually used to prospect the effect of certain tactics. To this end, based on a Markov chain model for predicting players' actions, Wang et.al.\cite{sportswhatif/tvcg/WangZDCXZZW20} design a visualization system to help table tennis analysts interactively simulate the game result of applying different player tactics.
% A detailed survey of prediction with visual analytics can be found in~\cite{cgf/LuGHGM17}. \hl{why not expanding this survey? This paragraph on prediction seems most relevant to us.}
Many existing work \cite{GuoDMKKLKZC19, XuGCGXQYC18} also applied deep learning models to compute the predictions for what-if and attribution. However, as most deep learning models are regarded as black-boxes, users are unclear why the deep learning model would produce certain results when doing what-if.

Causal analysis also can be used to accomplish what-if tasks by doing interventions on the causal graph~\cite{ijon/Shanmugam01}. Compared with the black-box deep learning models \cite{cnn, lstm}, causal analyses provide a better explainability since users can interpret how the predictions are generated by referring to the causal graph. Moreover, using causal analysis to conduct what-if can reduce the effect of data bias~\cite{ijon/Shanmugam01}. Despite the usefulness of causal analysis, few visualization researches have investigated applying causal analysis for interactively conducting what-if. In this paper, we seek to address this gap by designing tailored system designs and user controls for conducting what-if analyses on top of a causal graph.

% A set of useful applications can be done based on the causal graph. For example, users can use the causal graph to explain interesting phenomena, such as the different admission rates of males and females. According to the identified causal structure of multiple data dimensions, users can quantitatively measure the contribution of other dimensions to a specific dimension. They can further accomplish what-if tasks by doing interventions with the causal graph~\cite{ijon/Shanmugam01}.

% \fan{extend this section around causal applications of what-if analysis / attribution / decision making under uncertainty}
% \fan{discuss, there lacks an interactive visualization tool for helping users conduct the causal application tasks, visually, interactively, with trust / confidence. This paper fills the gaps by designing XXX features.}
\section{Informing the Design} \label{sec:interview}

This research is the result of a long-term collaboration with data analysts in a large technology company. The company collects a large amount of data about visitors of their online retail stores. By exploring the data, analysts hope to understand what kinds of behavior patterns or user characteristics are likely to influence the outcomes (e.g., product purchase, service subscriptions, and terminations).

The analysts currently use correlation models to characterize the relation between factors. However, correlation is a measure for describing the relevance between factors' values and cannot be used to answer questions like \textit{Does changing the value of A lead to the change of B}. Hence, the insights derived from their current correlation models were uncertain and obscure. These limitations motivated the analysts to apply causal models to investigate how the different factors interact with each other and how much each factor influences the outcomes.

In this section, we introduce an interview study with the analysts to collect their design needs that drive our system development.

% In this section, we introduce a real use case scenario we used in the interviews, the participants and interview process, and the design needs we collected.

% \subsection{Scenario}
% Consider a campaign use case scenario. A marketer Bob is establishing a campaign for promoting the consumption of a group of customers. Considering the constrains of budgets, Bob hopes to find efficient marketing channels for the campaign.  By applying correlation analysis on historical marketing data, Bob identifies a set of factors that are highly related with the consumption. Bob next needs to select channels from these factors and formulate an action plan. However, the selection becomes difficult as Bob is not sure about the true effect of each channel. For customers who receive sales emails and have consumption, it is hard to say all of their consumption are caused by receiving the emails, as part of them may already have shopping intention before receiving the emails. Struggling with the selection, Bob decides to select the top related factor as the channel. Bob next suffers from the formulation of action plans. Although Bob has the ability to formulate a comprehensive action plan, the performance is hard to predict. More importantly, it is hard to convince others that the action plan can work since Bob is not clear how the channel actually influences the consumption.

\subsection{Participants and Process}

We recruited five data analysts (one female, domain experience 4-8 years each) from the technology company, who were interested in causal analysis. Three of them were marketing analysts who were interested in adopting causal analysis for their customer profile and behavior data (P1-3). The other two were experts in causal analysis, who had more than three years of experience developing and applying causality-based models (P4-5).

We conducted two semi-structured interviews with the marketing analysts and causal experts, correspondingly. During each interview, we began by introducing the concept of causal analysis with the campaign use case scenario (described in Section~\ref{sec:introduction}). Then, we asked the participants to describe other causal analysis scenarios in their daily jobs, the tools they have used for conducting causal analyses, and the difficulties and needs with utilizing those tools. We encouraged analysts to share and describe the real challenges they have faced in different use cases. We also summarized the needs and conducted a follow-up interview with the two causal experts to verify the possibility of addressing these needs with causal analysis. For each interview, we had an experimenter responsible for taking notes and coding the transcripts.

\subsection{Design Needs}

Based on the interviews, we identified 7 key design needs across 3 major requirements. For the validity and the generalizability, each design need is mentioned by at least two interviewees.

% \new{\subsubsection*{R1. Support for Examining Causal Detection Result}
\begin{compactenum}[R1]
\item \textbf{Support for Examining Causal Detection Results}

\noindent The marketing analysts commented that an interface for \q{seeing the whole causal graph} can help understand the causal detection results and answer questions such as \q{What are the most related causal factors of an outcome?} However, the visualization tools they have used are not scalable for \textbf{the presentation of large causal graphs} (\textbf{N1 $|$ P1-3}). Moreover, the causal experts commented that the automatic causal discovery algorithms usually assigns different levels of uncertainty for each causal relation. Considering the reliability, the marketing analysts would like to focus on more convincing causal relations in their analyses. Hence, it is also important to \textbf{show the uncertainty of the detected causal relations} (\textbf{N2 $|$ P1-5}). The causal experts also emphasized that \textbf{inspecting the data quality with an interface} (\textbf{N3 $|$ P4-5}) is necessary for causal analysis since the causal detection usually requires certain assumptions in the data.

\item \textbf{Support for Identifying Influential Factors}

According to the marketing analysts, before purchases, users may receive multiple treatments simultaneously, such as discount e-mails and advertisements on social platforms. How to \q{correctly identify the contribution of multiple factors on a specific outcome} therefore becomes a major task for evaluating the existing marketing plans. Hence, the system should allow users to \textbf{quantitatively estimate the influence of each dimension} (\textbf{N4 $|$ P1-2, P4}). Moreover, rather than showing a numerical measure for each factor, analysts would like to know \q{how a marketing factor influence the outcomes} and \q{what are the intermediate variables from factors to outcomes.} This requires an \textbf{embedded visualization of both influential factors and causal information} (\textbf{N5 $|$ P1-2, P4}).

\item \textbf{Support for Making What-If Interactions}

The marketing analysts usually create a set of marketing plans to improve targeted outcomes. Although they could anticipate the effect of each plan based on their knowledge, the detailed change of each outcome is still unclear, which poses challenges for the decision-making process. Therefore, the marketing analysts expressed their needs of \textbf{simulating the marketing plans to see the possible effect} (\textbf{N6 $|$ P1-5}). P1 also emphasized that \q{notifying the side effect of a marketing plan} (e.g., some plans may increase the purchase in the next year but lower users' loyalties) would also be helpful for their works. This requires the system to \textbf{present the local as well as the global effect of an intervention} (\textbf{N7 $|$ P1-3}). 
\end{compactenum}

\section{Causal Modeling} \label{sec:model}

In this section, we introduce the background of causal modeling. We first provide a formal definition of causal graphs and then describe approaches for discovering the causal graph for a multi-dimensional tabular dataset. Finally, we introduce the uncertainty in automatic detected causal graphs.

\subsection{Causal Graph Definition}

The idea of using a DAG to represent the causality is from the structural causation model (SCM)\cite{ijon/Shanmugam01}. A causal graph is defined as $G = (V, E)$ where $V$ represents nodes and $E$ represents edges. Each node is a variable and each edge is a causal relation. For $X, Y \in V$, if $X$ is the parent of $Y$, then $X$ is said to be the cause of $Y$. If there is no edge between $X$ and $Y$, $X$ and $Y$ are independent when other variables are controlled, noted as $X \independent Y | Z, \exists Z \subseteq  V_{\backslash \{X, Y\}}$.
% \begin{equation}
% X \independent Y | Z, \exists Z \subseteq  V_{\backslash \{X, Y\}}
% \end{equation}
$V_{\backslash \{X, Y\}}$ represents all variables in $V$ except $X$ and $Y$.  For example, for a causal graph of three variables $<X, Y, Z>, $ the absence of the edge between $X$ and $Y$ (which are correlated according to the Pearson Index) means that $X$  and $Y$ are independent when conditioning on $Z$.

The independence between variables can be examined by conditional independence (CI) tests. The goal of CI tests is similar to controlled experiments, i.e., testing the true relation between variables by controlling other variables. It is popular to use partial correlation to do CI tests for numerical data. More descriptions of the CI tests for different types of data can be found in \cite{CItest}.
Following this definition, each causal graph $G$ can be mapped to a distribution $\widehat{P}$ over $V$. $\widehat{P}$ is a joint distribution of variables in $V$ and can be factorized as $\widehat{P} = \prod_{i=1}^{n}P(V_i|Pa(V_i))$
% \begin{equation}
% \widehat{P} = \prod_{i=1}^{n}P(V_i|Pa(V_i))
% \end{equation}
,where $n$ is the total number of nodes in $V$ and $Pa(V_i)$ is the set of parents of $V_i$. Therefore, a graph $\widehat{G}$ is equal to the true causal graph $G$ when its distribution $\widehat{P}$ is equal to the real data distribution $P$.

\begin{figure}[t]
	\centering
	\includegraphics[width=1\linewidth]{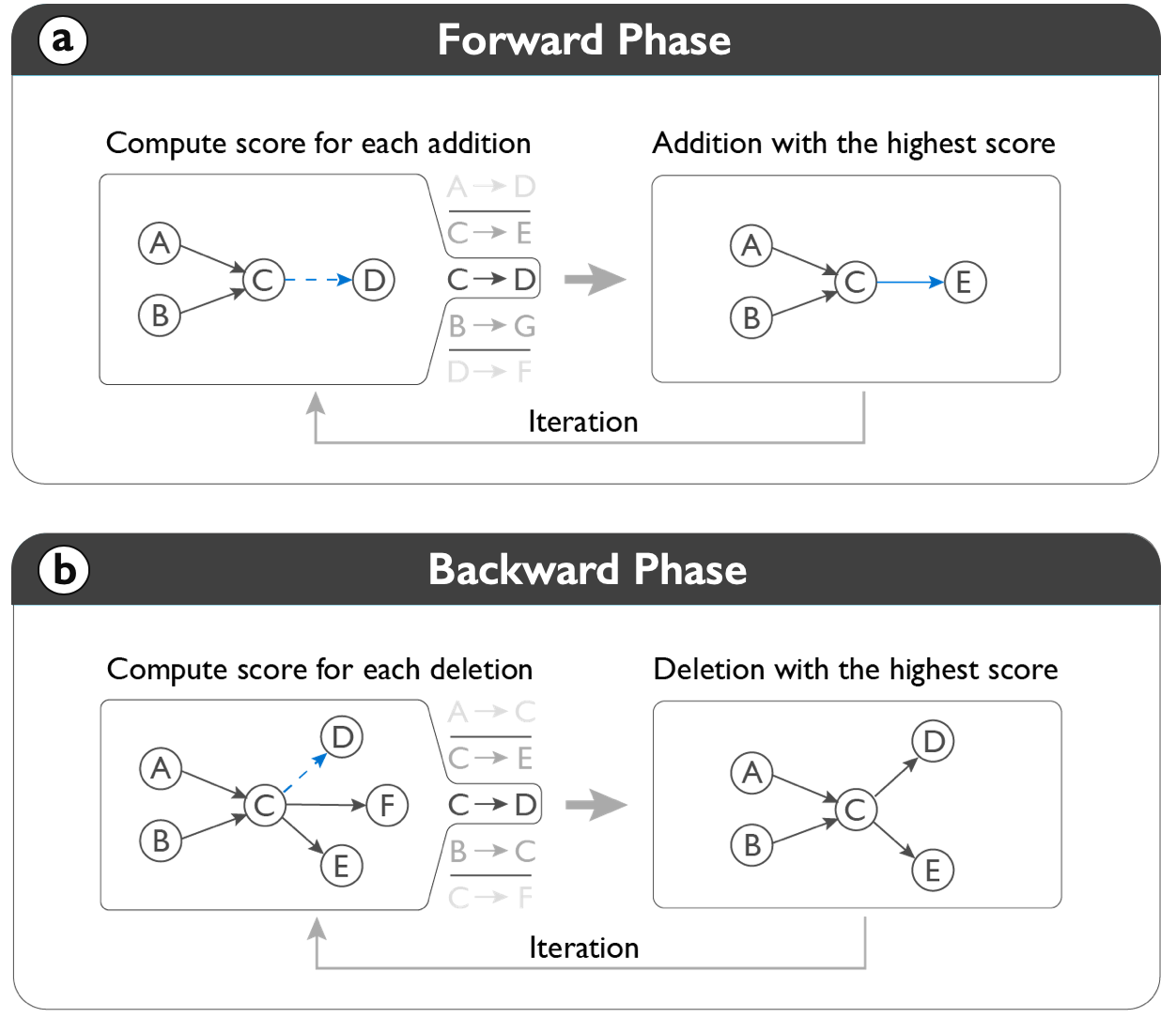}
    \caption{Explanation of the causal discovery method F-GES. The computation consists of a forward phase (a) and a backward phase (b). Forward phase: Given a causal graph, this phase will iteratively insert a new edge with the maximum score increase into the graph. Backward phase: Given a causal graph, this phase will iteratively delete an existing edge with the maximum score increase from the graph. Both the forward phase and the backward phase will be stopped when the score does not increase. }~\label{fges}
    \vspace{-7mm}
\end{figure}

\subsection{Causal Discovery}

According to the definition of the causal graph, the constraint-based methods are firstly proposed to detect causal graphs from tabular data. The algorithm will test the dependency of each pair of variables and for each pair there will be at most $(n-2)!$ numbers of conditions that need to be tested. Although researchers have proposed different approaches to reduce the number of required CI tests, doing one CI test is still very time-consuming. For example, the time complexity of partial correlation is $O(m^3)$, where $m$ is the number of data dimensions. Hence, the constraint-based methods, which are considered as precise but time-consuming, are not suitable for the big data scenario.

We apply the state-of-the-art F-GES\cite{ijdsa/RamseyGSG17} to detect the causal graph from big data. Here we briefly introduce the detection.
The detection contains two phases. The first phase is a forward phase (Fig.~\ref{fges}(a)).  Given a causal graph G, this phase iterates over every alternative one-edge addition. Fig.~\ref{fges}(a, left) shows that a new edge $C \rightarrow D$ is added to the existing $G$ and F-GES will compute a score for this addition. The score here is a measure of how well the causal graph can be used to fit the data distribution. A widely used score is Bayesian Information Criterion (BIC)\cite{bic1, bic2}:
\begin{equation}
BIC = \ln(n)k - 2\ln(L)
\end{equation}
where $n$ is the sample size, $k$ is the number of parameters, and $L = P(X|G)$ is the maximum likelihood. Hence, the score contains two parts, a penalty of the complexity of the causal graph structure and a fitness between the causal graph and the data samples.

An one-edge addition with the highest score improvement (add $C \rightarrow E$ in Fig.~\ref{fges}(a, right)) will be chose. The first phase iteratively conducts this one-edge addition until no more additions can improve the score. F-GES then proceeds to the backward phase (Fig.~\ref{fges}(b)). Backward phase is similar to forward phase except that one-edge addition is replaced by one-edge deletion (Fig.~\ref{fges}(b, left)). For each iteration, backward phase conducts the one-edge deletion with the highest score improvement (delete $C \rightarrow D$ in Fig.~\ref{fges}(b, right)). In this manner, F-GES obtains a causal graph that can fit the data distribution without much overfitting. Overall, the computation can be decomposed which allows parallel computation and the computation can be reused during the iteration. Hence, F-GES achieves a high scalability of dimensions.

% \subsection{Uncertainty}
Despite the effectiveness of causal discovery methods, the detected causal graphs often entail uncertainties of the causal link. As stated by~\cite{ijdsa/RamseyGSG17}, it is possible to introduce false-positive links into the causal graph. To estimate the uncertainty of a causal link $e$, we compute the BIC score difference of a causal graph with and without this link. i.e.,
\begin{equation}\label{uncertainty}
% Uncertainty(e) = \frac{1}{BIC(G) - BIC(G_{e})}
Uncertainty(e) = BIC(G) - BIC(G_{e})
\end{equation}
Here the uncertainty is computed after the backward phase of F-GES, which ensures that every edge in the causal graph meets $BIC(G) > BIC(G_{e})$. Hence, the uncertainty value is always positive.

\subsection{Intervention}
% \task{XIAO: move to here}
\label{intervention_process}

Intervention can be interpreted as an interaction of setting data dimensions to specific values and inspecting the effect. An intervention can be represented as a set of $<key, value>$ pairs. Keys represents the variables (e.g., \textit{weight}) and values represents the specific value of variables (e.g., 100kg). The result of an intervention is a set of distributions $\{d_1, d_2, ..., d_n\}$ where $d_i$ is the distribution of $V_i$. 
Here $d_i$ is interpreted as the possible distribution of $V_i$ when fixing variables' values according to the intervention.
Users can compare between $d^{1}_{i}$ (origin) and $d^{2}_{i}$ (after intervention) to see the effect.
For example, when trying to propose a new design of cars, users can set $<horsepower, 100>$ and obtain a set of distributions. 
They may find that $d^{2}_{mpg}$ is smaller than $d^{1}_{mpg}$ and reject this setting.
The intervention is accomplished by sampling over the causal graph. The detail is as follows.

% \subsubsection{Sampling over causal graph}

We first define a sample of the causal graph as $\{v_1, v_2, ..., v_n\}$ where $v_i$ is the value of $V_i$. According to the causal graph, $v_i$ can be sampled from its conditional probability distribution (CPD) $P(V_i| Parent(V_i))$. For example, when $v_{horsepower}$ is 100ps and $v_{displacement}$ is 2.0T, $v_{weight}$ can be obtained by sampling over its CPD $P(weight| horsepower=100, displacement=2)$. Particularly, the value of variables without any parents can be obtained by sampling over their probability distributions $P(V)$. Therefore a sample of the causal graph can be obtained by sampling variables following the topological order. When doing an intervention$<V_j, v_j>$, each variable's value can be sampled from $P(V_i|Parent(V_i), V_j = v_j)$. We can sample multiple times from the causal graph and compute a new distribution for each variable from the samples. These distributions are regarded as the intervention result.

\begin{figure}[t]
	\centering
	\includegraphics[width=1\linewidth]{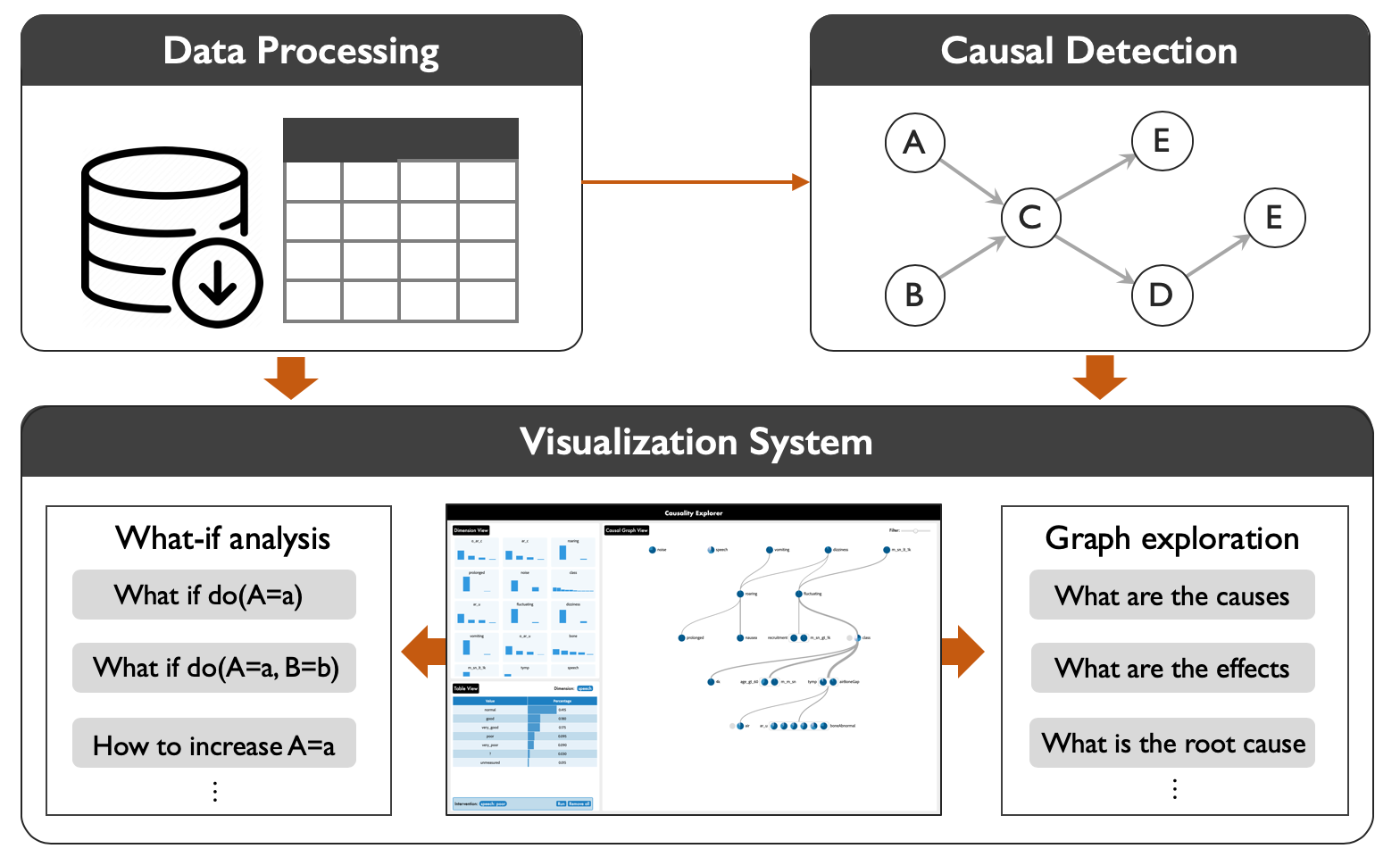}
    \caption{The system consists of three components, the data processing component for processing high-dimensional data, the causal detection component for computing the causal graph, and the visualization component for supporting the causal graph exploration and what-if analysis.}~\label{pipeline}
    \vspace{-7mm}
\end{figure}

\subsection{Attribution}
\label{attribution_process}
% \task{XIAO: move to here}

In marketing, attribution analysis is regarded as explaining why portfolios can create certain performance compared with the benchmark. Different attribution models, such as last-click attribution and probabilistic attribution, have been proposed for assigning credits. Causal graphs are also helpful for attribution analysis. A significant advantage of causal-based attribution is that the computation result is explainable, i.e., users can comprehend why a channel would be assigned certain credits. Conducting attribution with causal graphs is based on the operation of intervention. Given a dimension $V_t$ and one of its value $v_j^t$, we refer the attribution analysis as finding the effect of other variables on the proportion of $v_j^t$. To compute the effect, we will first identify variables that have paths to $V_t$, which referred as $S$, according to the causal graph. The rest variables are regarded as no causal effect. With $S$, we conduct the following computation process for every $v_j^i$
\begin{equation}
f(v_j^i) = Abs(P(v_j^t|do(V_i = v_j^i)) - P(v_j^t|do(V_i \neq v_j^i)))
\end{equation}
where $f(v_j^i)$ represents the effect of $v_j^i$ on $v_j^t$ and $P(v_j^t|do(X))$ represents the probability of $v_j^t$ when doing intervention $X$. Therefore, the effect of $V_i$ on $v_j^t$ can be computed as $Max(\{f(v_j^i)\})$.

\section{System Design} \label{sec:system}

Informed by the interview study, we iteratively designed \name for conducting exploratory causal analysis. During a 6-month iteration, multiple prototypes were designed and tested with domain practitioners before reaching the final system.
In this section, we describe how we designed the \name system based on the user needs gathered from the interview study. For simplicity, we use an audiology dataset \cite{audiology} (200 rows, 24 categorical attributes) from the UCI repository to illustrate the main functionalities and the causal graph layout designed for high-dimensional data.

\subsection{Overview and Workflow}

The \name system consists of two major interface components for addressing the user needs (Fig.~\ref{pipeline}): a graph view for exploring the causal relations (R1) and a what-if analysis view for simulating a specified interventions (R2) or for detecting attributing factors for a specified goal (R3). The main workflow of this system is as follows. Users will explore the causal graph first and learn the convincing causal mechanism embedded in the data. According to users' prior knowledge or domain-specific requirements, they may focus on the improvement of specific data dimensions and utilize the attribution component to find a set of options that are helpful for the improvement. Finally, users will test over the options with the what-if component and make decisions according to the test result.

The rest of this section will describe the design of each component and introduce our design process and rationales. We also provide implementation details at the end of this section.

\subsection{Causal Graph Visualization}

% According to Guo et.al \cite{tvcg/GuoHL15}, the most intuitive channels for visualizing uncertainties with line marks are \textit{lightness}, \textit{fuzziness}, \textit{transparency}, and \textit{grain}. 
% Considering nodes and edges as the two main components of a causal graph, we raise two design criteria respectively.

As stated by R1, a causal analysis usually starts with an exploration of the causal graph. To this end, we propose a novel scalable causal graph visualization to support the causal analysis of high-dimensional data.

\subsubsection{Encoding of Nodes and Links}

As shown in Fig.~\ref{system}(a), in the graph visualization, each dimension is represented by a piechart (Fig.~\ref{system}(d)) where each sector encodes the proportion of a dimension value. This can help users learn the characteristic of each dimension and provide guidance for exploration and validation. For example, it can help users quickly filter out dimensions that most instances share the same value.

Links indicate the causal relation and the direction is consistently from the upper node to the lower node. For example, the connection between \textit{roaring} and \textit{nausea} (Fig.~\ref{system}(e)) means that \textit{roaring} is the cause of \textit{nausea}. The uncertainty of a causal link, which is computed in Sec.~\ref{uncertainty}, is encoded by the degree of thickness (Fig.~\ref{system}(f)) where a thicker link represents a more confident relation. As stated by Guo et.al\cite{tvcg/GuoHL15}, different visual channels, such as color, lightness, and transparency, are available for encoding the uncertainty of links. Regarding the uncertainty as the most important feature of a link, we decide to use the thickness channel, one of the most effective channels of line, as the visual representation. Users can double click on a node and the causality subgraph of this node will be displayed (Fig.~\ref{system}(h)).

\subsubsection{Graph Layout}

The position of each node is determined based on its related causality, i.e., the vertical position of a node is higher than each of its child nodes in the causal graph. With this layout, users can quickly identify the causal direction and the related causal factors of a node. This layout is formulated according to the discussion with experts. Based on the discussion, two design criteria are proposed for locally and globally explore the causal graph respectively.

\begin{compactenum}
\item \minisection{The direction of each link should be explicit} When locally exploring a causal graph, the most important task is to find the causes of a specific node. Emphasizing the direction information is helpful for the cause identification.
\item \minisection{The role of each node should be clear} When globally exploring a causal graph, identifying two types of nodes, the \textit{root} with 0 in-degree and the \textit{leaf} with 0 out-degree, is helpful for perceiving and diagnosing the graph. The two types of nodes can seem as an analogy to the input and output of a causal graph.
\end{compactenum}

% After satisfying the two primary criteria, a legible causal graph layout should also satisfy common criteria for graph visualizations, like the number of link crossing should be reduced and the layout should be scalable. To fulfill the criteria, we propose a two step heuristic causal graph layout method as follows.

Here, we describe how to generate a legible causal graph that can satisfy the two criteria. We adapt existing layered graph layouts \cite{battista1998graph} and techniques for reducing edge-crossings \cite{algorithmica/EadesW94} to the causal graph visualization. Although the layered graph layout has been applied in many existing applications, adapting these approaches to the visualization of a large causal graph still encounter multiple challenges, such as the cross-layer causal links and the aggregation of causal structures. The details of generating a tailored layer graph for visualizing a large causal graph are as follows.

\begin{figure}[!htb]
	\centering
	\includegraphics[width=1\linewidth]{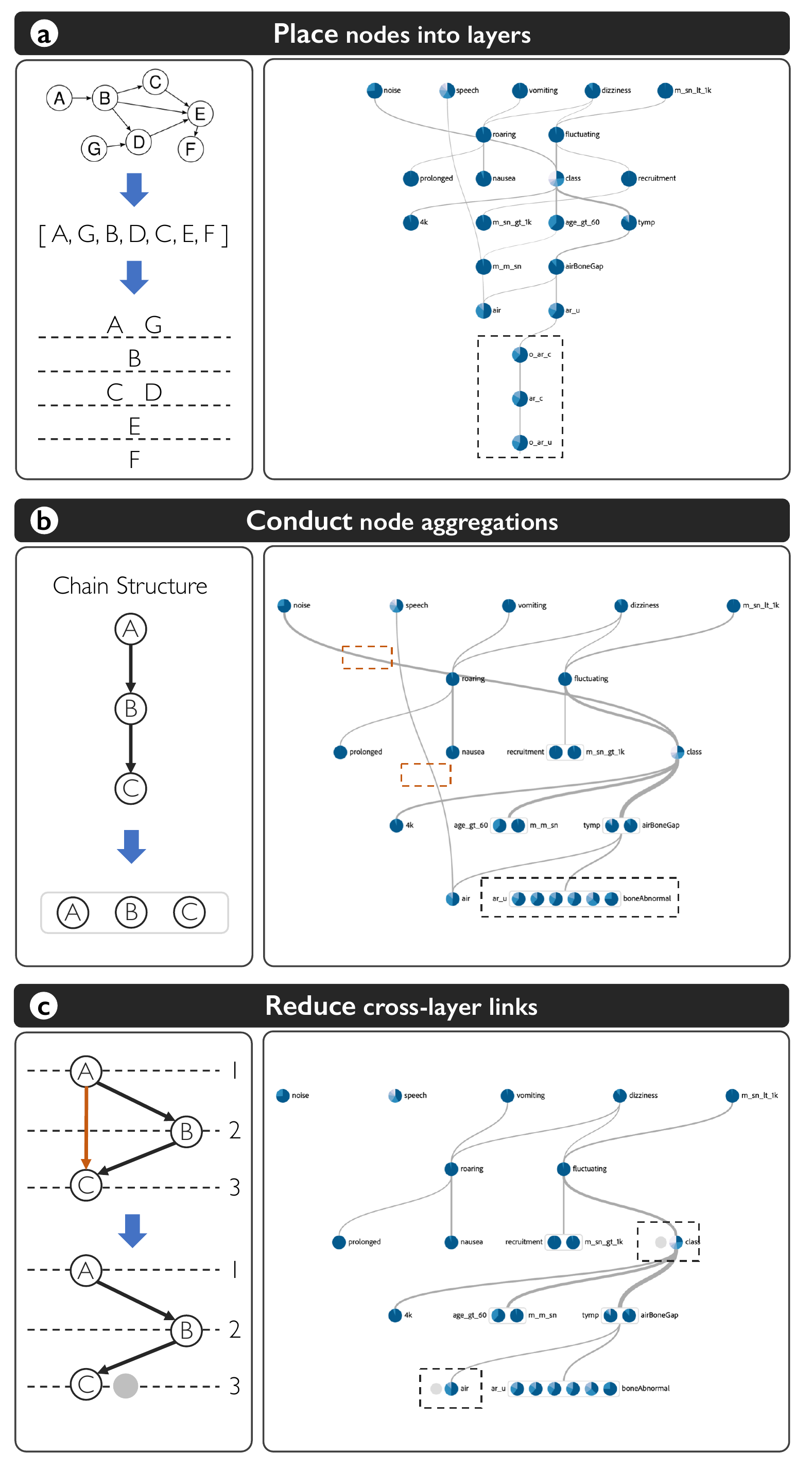}
    \caption{Process of producing a legible layout for a large causal graph using the audiology dataset \cite{audiology} (200 rows, 24 categorical attributes). (a) Divide nodes into layers according to the topological order to ensure the readability of the causal directions. (b) Find the chain structures and aggregate the nodes in the chains to increase the visual scalability. (c) Render cross-layer links as glyphs to reduce the visual clutter.}~\label{graph_layout}
    \vspace{-8mm}
\end{figure}

\subsubsection*{Step 1: Layout Nodes by the Topological Order}

This step is to fulfill the first criteria. The direction of links is usually indicated by arrows in DAG. This encoding, however, can create severe visual clutter for a large causal graph.This step is to place nodes into different layers where all the causes of a node are from the precedent layers. The idea is to use the most efficient visual channel (positions) to encode the most important information (directions). This problem can be addressed by finding a topological order of nodes.
The topological order is commonly seen in a dependency graph. In this order, each node is given after all its dependent nodes (Fig.~\ref{graph_layout}(a, left)). The topological order can be acquired for every DAG \cite{topological_sort} and we use this order to form the layer of each node as
\[Layer(N) = Max(\{ Layer(N_i)  |  N_i\in C(N)\}) + 1\]
where $N$ represents a node and $C(N)$ represents all causes of a node N. The layer of each root node is set as 0. Each node (Fig.~\ref{graph_layout}(a, right)) is placed under all its dependent nodes (i.e. its causes) and the causal direction is from up to down. With this layout, we can find that there are 5 root causes (the top nodes of Fig.~\ref{graph_layout}(a)) in the audiology dataset. 

\begin{figure}[b]
	\centering
	\vspace{-4mm}
	\includegraphics[width=1\linewidth]{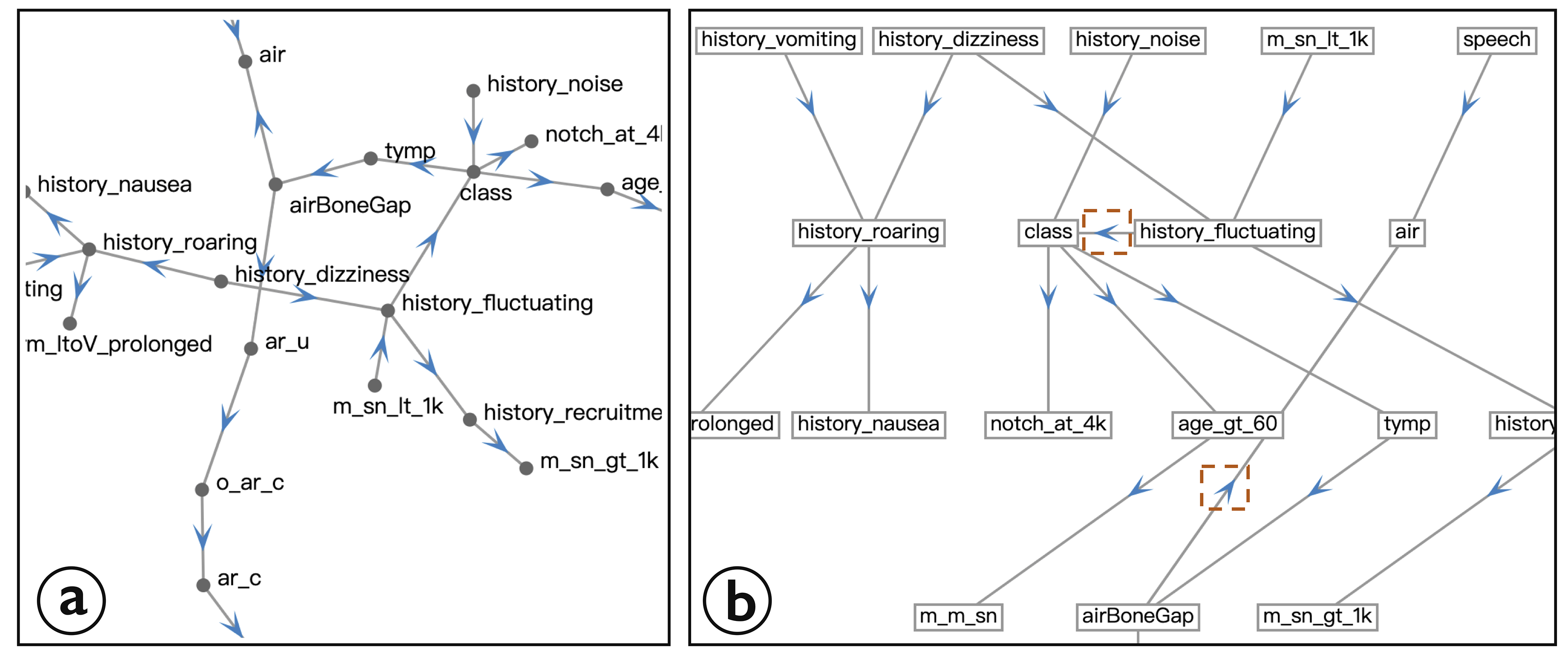}
    \caption{The alternatives of causal graph layout. (a) A force-directed layout. Although it can show nodes and links in a scalable manner, the readability of the causal link is low. (b) A spanning-tree layout. Most directions of the links are consistent. However, a few links (highlighted) with different directions may mislead users.}~\label{layout_alternative}
    \vspace{-2mm}
\end{figure}

\minisection{Node Aggregation}
The result of step 1 may face important scalability issues, i.e., the number of layers could be very large and users cannot inspect the whole causal graph at a glance. To address this issue, we extract a special causal structure, causal chain (Fig.~\ref{graph_layout}(a, black)), from the graph. There are three main causal structures in a causal graph \cite{sgs/spirtes1991}. Among the three structures, the chain structure (e.g., ``A'' to ``B'' to ``C'') is considered to be semantically-simple while having significant effect on the number of layers. For example, a causal chain with length $L$ would require $L$ layers to present this structure. Therefore, we transform the chain structures to aggregated nodes. 
% , which represents a cascading relation of causality,
%Each aggregated node is linked to the causes of its head (the first node of a chain) and the result of its tail (the last node of a chain). %
The process can be found in Fig.~\ref{graph_layout}(b, left) and the corresponding layout can be found in Fig.~\ref{graph_layout}(b, right).
Note that we only aggregate chain structures that have no links to other nodes out of the chains. 
% We discuss the level of aggregation in Section\ref{}.

\minisection{Cross-Layer Links}
This layout also leads to cross-layer links which can create visual clutter (Fig.~\ref{graph_layout}(b, orange)). The cross-layer links refer to links that connect nodes across more than one layer. For example, the link \textit{A} to \textit{C} (Fig.~\ref{graph_layout}(c, left)) is a cross-layer link as \textit{A} is in layer 1 while \textit{C} is in layer 3. We have found two options to address this issue. The first one is to turn the cross-layer links to orthogonal links to avoid the clutter. This is useful when the number of cross-layer links is limited. However, when dealing with a complex causal graph, multiple orthogonal links may intersect with each other and causes difficulties for the link perception. 
The other option is to hide the cross-layer links and use glyphs to encode the cross-layer causes. For each node, if there is a cross-layer link connected to this node, we will place a glyph by this node to encode the hidden causes. As shown in Fig.~\ref{graph_layout}(c, left), for the link \textit{A} to \textit{C}, we hide this link and place a glyph near the node \textit{C} to represent that there is a cross-layer cause. Users can hover on node \textit{C} to see the detail. Considering the scalability, we adopt the second option in our system. The layout after this step can be found in Fig.~\ref{graph_layout}(c, right). The current design uses the number of glyphs to encode the number of hidden cross-layer causes. This is to keep users aware of how many cross-layer causes they need to search when hovering over the node.
 
\subsubsection*{Step 2: Refine Layout by the Role of Nodes}

This step is to fulfill the second criteria. After step 1, all the root nodes, i.e., the node without any linked causes in the causal graph, are placed in the first layer. The leaf nodes, however, are scattered in different layers and hard to identify. To highlight the leaf node, we place these nodes on the left side of layers. As shown in Fig.~\ref{system}(g), \textit{prolonged} is a node without any out-degree and therefore is placed at the left. We did not choose to use popular highlight techniques like colors and sizes to ensure a consistent encoding (position) of leaf nodes and root nodes. After setting the position of these two types of nodes, the layout will be refined to reduce the number of link crossings. Reducing link crossings of bipartite graphs is NP-hard\cite{algorithmica/EadesW94}. Here we use a greedy approach to reduce link crossings under the constraint of placing leaf nodes to the left side of each layer.

% \task{XIAO: add alternative design discussion to graph. Objective pros and cons plus feedback from experts / design partners.}

\subsubsection{Design Alternatives}

The design is an iterative process and certain design alternatives are produced. Regarding the scalability as the major issue, we first consider the application of the force-directed layout. Due to its efficiency of reducing visual clutter and preserving community information, force-directed layout is widely adopted for visualizing large graphs and has also been used to visualize causal graphs \cite{tvcg/WangM16}. We apply this layout on the marketing dataset and present it to our experts (Fig.~\ref{layout_alternative}(a)). However, the experts commented that the causal directions are hard to perceive, as there are numbers of arrows in the graph. It is also hard to track the causal path between variables. Recognizing this problem, we consider the readability of causal links as the first priority issue and implement the sequential layout (Fig.~\ref{layout_alternative}(b)) based on a spanning tree algorithm\cite{ieeevast/WangM17}. The experts appreciate this layout. However, when applying it to a large causal graph, various inconsistent causal directions are identified. Although most links have a top-to-down causal direction, a few links within the same layer have different directions. It is hard for experts to quickly notice this inconsistency. According to users' comments, we further design the current layout to address the readability and the scalability issue for better accomplishing causal tasks.
During our design process, the largest causal graph that we have explored with this layout contains 186 edges and 100 nodes, which is already considered as a very large graph by our domain experts. Hence, the experts appreciated this layout and regarded it as an applicable solution.

% In our description, “scalable” refers to the number of nodes and edges of a causal graph that can be clearly visualized. The datasets and models in previous work only produce relatively small causal graphs (e.g., at most 13 nodes in [48]). During our iterative design process, the largest causal graph our system has produced for the domain practitioners contained 186 edges and 100 nodes. This graph has far exceeded the capability of existing solutions and was considered “a more realistic scenario” by the practitioners. We have added this clarification in Sec. 5.2.3.

% The “scalable” refers to the number of nodes and edges that can be visualized in the causal graph. The causal graphs in previous works are usually small (e.g., at most 13 nodes in the cases of VAST 15). Our causal graph layout can handle a larger number of nodes and edges. The largest causal graph that we have explored during the iterative design process contains 186 edges and 100 nodes, which is already considered as a very large graph by our domain experts. Therefore, our causal graph layout is more applicable than previous methods.

\begin{figure}[t]
	\centering
	\includegraphics[width=1\linewidth]{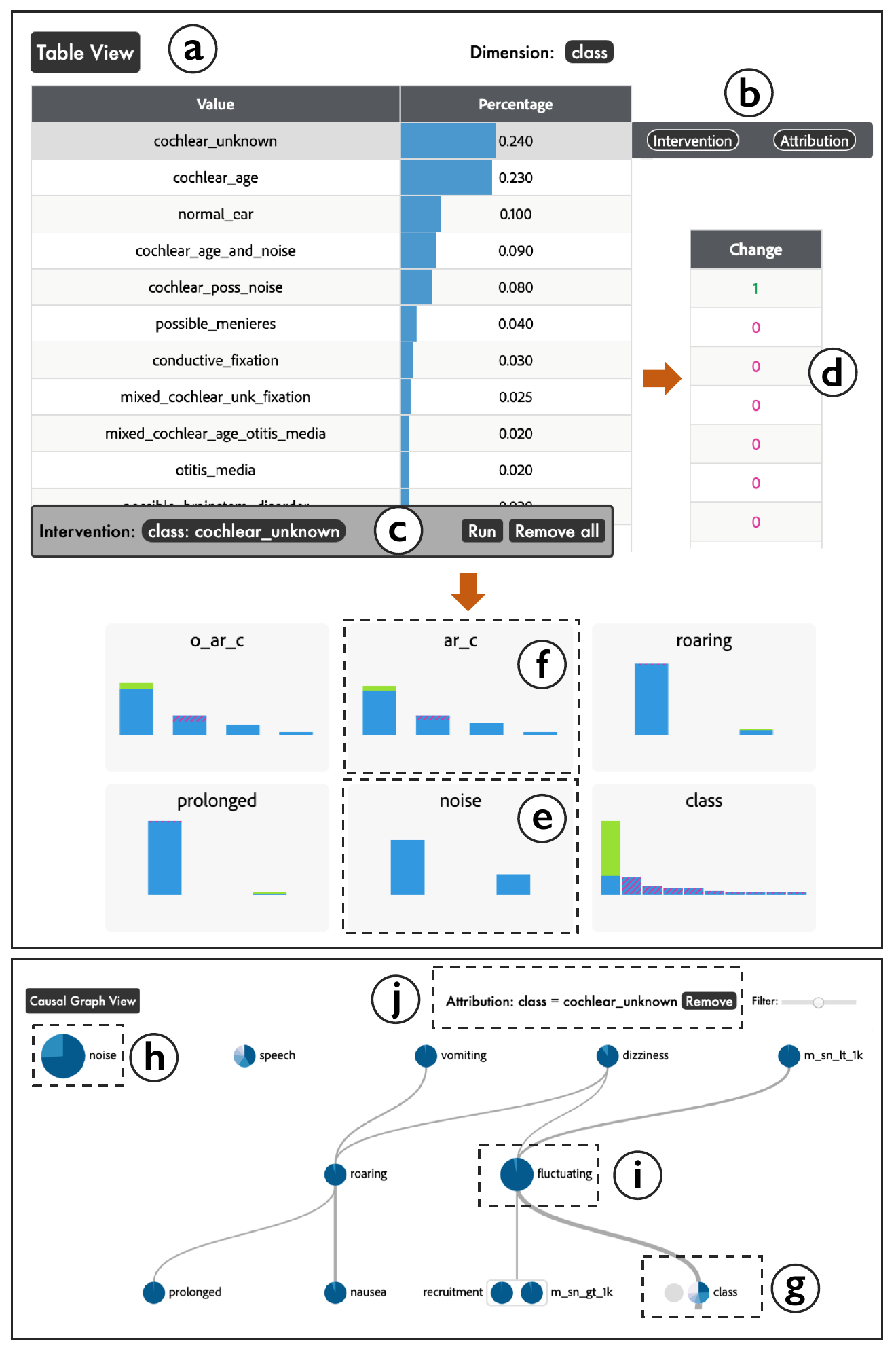}
    \caption{Interfaces for conducting intervention and attribution. Users can turn to the table view (a) and hover on a row to open a control panel (b). Clicking the intervention button (b) can set the dimension to a specific value (c) and the result of intervention will be updated in the table view (d) and the dimension view (e, f). Users can also click the attribution button (b) to find influential channels on this dimension value. The attribution result of the specified value (g) will be presented in the causal graph (h, i). A larger size of nodes represents a larger influence. Users can click the button (j) to remove the attribution result.}~\label{whatif_interface}
    \vspace{-8mm}
\end{figure}

\subsection{What-If Analysis}

Users can accomplish interventions and attributions by interacting with the Dimension view and the Table view. In the Dimension view (Fig.~\ref{system}(b)), each histogram represents the distribution of a dimension. We use the bar height to encode the proportion of a value and arrange the x coordinate of each bar according to the descending order of bar heights. Due to the limited space, the histogram shows the top 10 proportions when a dimension contains numerous values. Users can click on a histogram and the detail of the dimension will be shown in the Table view (Fig.~\ref{system}(c)). Each row shows the name and the proportion of a dimension value. When hovering on a row, a control panel is provided to help users establish the intervention and attribution (Fig.~\ref{whatif_interface}(a)).

\subsubsection{Intervention}

Users can click on the intervention button (Fig.~\ref{whatif_interface}(b)) to fix the value of a dimension. For example, Fig.~\ref{whatif_interface}(c) shows that users are setting the $class$ to a specific value $cochlear\_unknown$ ($do(X = x_1)$) for all the instances. Users can iteratively fix the value of dimensions ($do(X = x_1, Y = y_1)$) and the intervention setting will be stored in a panel (Fig.~\ref{whatif_interface}(c)). By clicking on the run button (Fig.~\ref{whatif_interface}(c)), the backend will compute the effect of this intervention on all the other dimensions. According to the computation process (Sec.\ref{intervention_process}), the effect on a dimension is represented by an estimated distribution. The estimated distribution will be updated in the table (Fig.~\ref{whatif_interface}(d)) and the dimension view (Fig.~\ref{whatif_interface}(e, f)). 

We propose a design named \textbf{diff bar chart} to help users more easily compare between the original distribution and the estimated distribution of multiple dimensions. As shown in Fig.~\ref{whatif_interface}(f), the original proportion is encoded by the blue bar. The increased proportion is encoded by the green bar and the decreased proportion is encoded by the red bar with a texture. We use the texture to emphasize that the cover region ``disappear'' after the intervention, which is more intuitive and appreciated by users. Fig.~\ref{whatif_interface}(f) shows that with the intervention, the proportion of ${ar\_c}$ has been changed while the proportion of \textit{noise} (Fig.~\ref{whatif_interface}(e)) is consistent. Users can inspect over the diff bar charts to obtain an overview of the effect of an intervention. Moreover, users can click on a diff bar chart to see the detail of a dimension in the Table view.  Users can click the remove button (Fig.~\ref{whatif_interface}(c)) to clean the result.

\begin{figure}[t]
	\centering
	\includegraphics[width=1\linewidth]{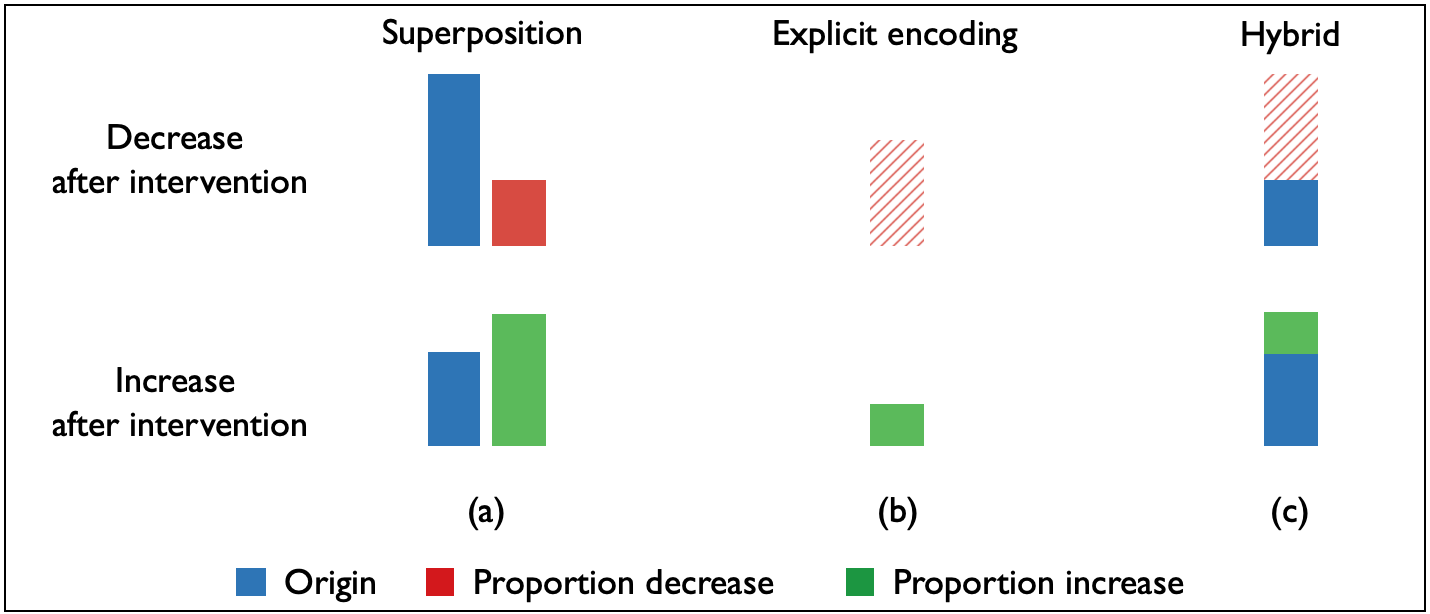}
    \caption{Design alternatives of visualizing the original distribution and the estimated distribution for the visual comparison.}~\label{diff_alternative}
    \vspace{-8mm}
\end{figure}

\subsubsection{Attribution}

Users can click on the attribution button (Fig.~\ref{whatif_interface}(b))  to identify the contribution of each dimension to the clicked dimension value. For example, when clicking the attribution button of \textit{class = $cochlear\_unknown$}, the causal graph will be updated accordingly to show the result of current attribution. As stated by Sec.~\ref{attribution_process}, the contribution is represented as a percentage value. We use the size of a causal node to encode its contribution where a larger node represents a larger contribution. In this case,  \textit{noise} (Fig.~\ref{whatif_interface}(h)) contributes most to the dimension value  \textit{class = $cochlear\_unknown$} (Fig.~\ref{whatif_interface}(g)) while \textit{fluctuating} (Fig.~\ref{whatif_interface}(i)) is the second largest dimension. This means that users can try to change the value of {noise} and {fluctuating} if their target is to change the proportion of  \textit{class = $cochlear\_unknown$}. Users can click the remove button (Fig.~\ref{whatif_interface}(j)) to clean the attribution result.

\subsubsection{Design Alternatives}
During the design process, we have identified several alternatives of the diff bar chart. There are four basic techniques for conducting visual comparison \cite{visualcomparison}, i.e., juxtaposition, superposition, explicit encoding, and animation. Juxtaposition places bar charts of the original distribution and the estimated distribution separately, which is not effective as the corresponding bars of the same dimension value are apart from each other. Animation is widely used to show the transition of data changes. However, since our comparison involves a set of dimensions and values, it is hard for users to track the concurrent change of multiple visual elements. Therefore, we explore the rest design space and propose three different designs based on superposition, explicit encoding, and hybrid, respectively. 
For the superposition (Fig.~\ref{diff_alternative}(a)), we place the original bar and the estimated bar side by side for the comparison. Although it is intuitive, this would require a much larger visual space than the histogram since a dimension usually contains various values. For the explicit encoding (Fig.~\ref{diff_alternative}(b)), we present the computed difference between the original and estimated proportion. However, users commented that the absolute value before and after the intervention is also important. For example, the purchase rate increasing from $10\%$ to $15\%$ is much better and harder than from $5\%$ to $10\%$. We summarize users' comments and propose a hybrid design (Fig.~\ref{diff_alternative}(c)). Users can observe the absolute bar height and the difference concurrently.

%1. Variables that have changed significantly under interventions should be highlighted.

% \subsection{Causal Attribution}

% Here attribution is regarded as a process of finding possible options to improve a variable (referred as target). For example, users may be interested in increasing the \textit{mpg} to reduce the oil consumption. All variables on the causal paths of \textit{mpg}, such as \textit{weight}, \textit{horsepower}, and \textit{origin}, could be possible options of interventions. The most primitive idea is to iterate over all these options, which is time-consuming and labor-intensive. Therefore, we provide the causal attribution to help users identify valuable options for interventions. Here we refer ``valuable'' as doing interventions on a variable can have significant effect on the target.

% \task{NEED MORE DETAILS}
%1. Ancestors of the target should be clear.
% ********************************************************************************
% \begin{figure}[t!]
% \centering
% \includegraphics[width=0.95\columnwidth]{figures/design}
% \caption{The final design: (a) XXX.}
% \label{fig:design}
% \vspace{-3mm}
% \end{figure}
% ********************************************************************************
\section{Evaluation} \label{sec:evaluation}

In this section, we report two case studies in different domains to investigate the applicability of our system. We also conducted interviews with domain experts to discuss the usability and limitations.

\subsection{Case Study I: Education}

In the area of education analysis, an important topic is about analyzing the school dropout \cite{dropout1, dropout2, dropout3}. For a university, there will be cases of school dropouts every year. The analysts aim to find out reasons for the school dropout and identify possible improvements to the school system to reduce the dropout rate. We invited two analysts (an advisor of the school department and a Ph.D. student of the College of Education) to conduct this case study.

\subsubsection{Dataset}

The analysts provided a dataset of 3,500 students from a college. The dataset includes students' personal status and their course grades. All the provided data has been anonymized. The personal information includes $Gender$, $Region$, $Political\ Status$, $Graduated\ Highschool$, $Major$, and $Student\ Status$. The course grades are provided as a list in which each row contains the course name, the corresponding credit, the student id, and the grade of the student. For each student, we aggregate the course grades into two dimensions, $GPA$ and $Fail$. $GPA$ is a categorical data which categorizes students' grades into four levels according to the 4.0 scales. $Fail$ is a binary dimension which represents whether a student has records of failing an exam. 

\subsubsection{Process}

The analysts first focused on the causal graph to inspect the detected causal relations (R1). By exploring the causal graph, the analysts found two nodes on top of the graph, i.e., $Gender$ and $Region$. The experts agreed with the result as these two dimensions apparently cannot be influenced by other dimensions. The analysts then iteratively validated each link's truthfulness according to their knowledge. The link $Region$$\rightarrow$$Graduated\ Highschool$ first attracted the analysts' attention. The thickness of the link indicated that the model is confident that $Region$ is the cause of $Graduated\ Highschool$. The analysts commented that students usually graduate from their local high school and it was glad to see that this straightforward causal relation is identified, which significantly increased their confidence in the detection. The correctness of other links was also verified in later stages, such as $Gender$$\rightarrow$$Major$, $Region$$\rightarrow$$Major$.

Finally, the analysts examined the causal factors of $Student\ Status$. $Fail$ was connected to $Student\ Status$ and beside the node of $Student\ Status$, there were two glyphs representing two cross-layer causes. The analysts hovered on the node of $Student\ Status$ and found that the cross-layer causes were $GPA$ and $Major$. It was expected that $GPA$ and $Fail$ would have links pointed to $Student\ Status$ as the most direct reason for a student's dropout is that he/she is not able to finish studies. However, the link of $Major$$\rightarrow$$Student\ Status$ was unexpected. The expert commented that this represented that certain majors might have inappropriate settings or disciplines which therefore affected students' dropout.

To find possible ways of reducing the dropout rate, the analysts selected the dimension of $Student\ Status$ in the dimension view and set the attribution as $Student\ Status = dropout$ in the table view (R2). After setting the attribution, the analysts observed that the size of nodes in the causal graph changed and the largest node was $Fail$. However, this was the dimension that cannot be directly intervened and therefore the analysts decided to try interventions of $Major$ (The second largest node). The college had 12 different majors while four of them account for more than 80\% of students. The analysts first iteratively set the four main majors as interventions and found all of them can lead to a decrease of the dropout rate (R3). In addition to the four majors, the rest of the majors are mainly collaborative projects except for a special class, which was established to recruit students with high entrance marks and was managed differently compared with regular majors. Setting the major to this class, the analysts found that the dropout rate had a significant increase from $1.3\%$ to $7\%$ (R3). The analysts hypothesized that students in this class may struggle with great pressure due to the sense of competition. Applying additional psychological counseling to this set of students should help reduce the dropout rate.

\subsection{Case Study II: Digital Marketing}

To understand the applicability and usefulness of \name in digital marketing scenarios, we conducted a case study with the three marketing analysts (P1-3), who participated in our needfinding interviews and our system prototyping iterations. The case study lasted about three months through bi-weekly meetings, consisting of requirement discussions, data preparation, and data exploration.

\subsubsection{Dataset}

The marketing analysts provided a real data sample of the visit logs of an online retail store. Each row in the log represents a visit and the columns record different dimensions about the visit, such as the device type and location of the visitor, the referral channel and landing page of the visit, and if a purchase was made during the visit. The data contained 10,000 visits sampled by a time window and 32 dimensions. The analysts categorized the data dimensions into three types:
\begin{compactitem}
\item \textbf{Outcomes.} dimensions that are considered as success metrics in the analyses, such as the number of purchase orders or the click-through rate of ads.
\item \textbf{Interventions.} dimensions that can be directly managed by marketing tactics. For example, marketers can prioritize the targeted locations of campaigns or adjust the investment across different referral channels for their websites.
\item \textbf{Observations.} dimensions that cannot be directly changed by marketers, such as visitors' browser or device types, or their internet connections (e.g., $Lan/Wifi$ or $Mobile$).
\end{compactitem}

\subsubsection{Process}

After loading the dataset and the causal graph, the analysts decided to start by reviewing the graph nodes and links to check if the data were correctly visualized. The analysts carefully inspected the value distribution of each dimension by exploring the Graph View and Table View (R1). 
They found that while some of the nodes have a balanced distribution of the values, many were dominated by a population one (e.g. $Referral\ Channel$, $Browser\ Type$, and $Language$). Also, two of the nodes had only one single value ($JavaScript version$ and $Device\ ID$). \q{The piecharts around the nodes are extremely helpful,} P1 commented, \q{in a few minutes I already see several data ingestion problems that we need to report to the data engineering team.}

After confirming that the data is correct, the analysts started to explore and discuss the graph links (R1), which showed the causal relations between the nodes. All the analysts found the link easy to understand. P1 commented that \q{I like the top-down layout. It is very easy to keep track of what caused what.} After a short exploration, the analysts identified several causal relations that they were expected to see, such as $Country$$\rightarrow$$City$ and $Referral\ Channel$$\rightarrow$$Landing\ Page$.
They also observed that the links for representing these causal relations are all thick lines, which indicated a low uncertainty and further confirmed their assumptions on the data. P1 added that \q{these strong lines look very intuitive to me. I immediately knew they are the reliable results that I need to pay attention to.
Several causal relations were new and unexpected, such as $Referral\ Channel$$\rightarrow$$Number\ of\ Searches$ and $Browser\ Type$$\rightarrow$$Operating\ System$. \q{It seems there are far more causalities in the data than I know about} P3 commented excitedly. However, P3 requested to gather more data to verify these findings. \q{The links show a relatively high level of uncertainty compared to the rest of the graph,} he explained.}

% We have added additional details in the case study (Sec. 6.2.2) to demonstrate how the uncertainty facilitates users’ analyses. Overall, we found the uncertainty information critical to analysts’ decision making process. For example, it helps them identify and focus on causal relations that are certain and important, and ignore those that are less certain and less important. We also observed that analysts sometimes also paid attention to causal relations that are less certain but important to make sure they were not missing any potential high-value signals.

To narrow down the analysis, the analysts clicked on the $Purchase$ node, which represents the outcome in the analysis, and the graph was reduced to 7 nodes that have causal relations with the outcome. The analysts hovered on $Purchase$ node and the Table View showed the number of visits that led to at least one purchase order and those that led to zero. The analysts were surprised that the purchase rate was higher than usual during the time window of the sample, and clicked on the attribution button to analyze how much influence each dimension had on $Purchase = true$ (R2).

From the attribution results, $Login\ Status$ had the largest influence while $Landing\ Page$ and $Referral\ Channel$ had a similar but smaller influence. \q{These factors are exactly what I was thinking about,} P1 commented, \q{we can probably adjust the referrals or land more traffic to a certain page, but it would be hard to make people register or login.} P2 agreed and proposed to perform what-if analysis on the $Landing\ Page$ since \q{it is very easy to verify through A/B testing}.
The analysts one-by-one selected the 10 most popular landing pages and reviewed the changes to purchase rate (R3). Compared to $Homepage$, they identified three alternatives that had a positive influence on purchase rate, including $Product\ Search$, $Product\ Category$, and $Purchase\ History$. They decided to formulate A/B testings to further verify the results and share with their product managers.

At the end of the analysis, P1 commented that \q{this tool is very flexible to use and the graph provides a clear picture about what is important to purchase and what are not.} P2 suggested testing the causal model with data from a larger time window and evaluate the accuracy against A/B testing results. P3 requested a function to support the comparison of multiple what-if simulation results.

% P2 added that \q{A causal model has corrections for the compound effects so the results are more trustworthy.}
\section{Discussion} \label{sec:discussion}

The case studies suggest that \name is helpful for accomplishing tasks of causation exploration and what-if analysis. The analysts are able to identify the main causal factors of important dimensions and clear causal pathways from the causal graph. The interaction is also useful for testing the effect of different interventions.
The analysts were excited about this tool. \q{The relation provided by the causal graph is really clear.} They liked the layout of placing nodes in layers which presented the causal direction explicitly. The animation of the causal link is also appreciated, as it is intuitive and aesthetic. The usability and the effectiveness of the what-if analysis is approved by all the analysts. One analyst commented that he can formulate quantitative evidence of the effect of certain actions when giving a report.

The analysts also provided several useful suggestions for improving the system. First, the dimension view can be further improved by adding richer interactions, such as deleting and merging dimensions, so that users can immediately resolve minor data preparation issues without leaving the system or losing the already performed analyses. The analysts also commented that the what-if interactions can be improved by tracking the history of trials instead of only showing the latest results, so that they can easily compare the effects of different action plans. Adding a new dedicated panel for comparison tasks was also requested.

% \new{The analysts also provided several suggestions for improving the system. \q{The dimension view should provide more interactions for controlling the data dimensions.} The analysts preferred to accomplish feature engineering tasks such as deleting and merging dimensions with the system so that they can immediately improve the dataset when they notify several data quality issues from the causal graph. The analysts also commented that the what-if interactions can be improved by recording the history. They felt difficult comparing the effectiveness of different what-if interactions since the system only presented the latest what-if result. Adding a new panel for comparison can solve this problem.}

From the discussion with experts, we have identified a set of implications and summarize it as follows.

\textbf{Model Explainability.} The first implication is about the need of explaining the detected causal links with visualizations. From the case studies, we observe that users commonly ask questions about why there is a link between the two nodes. Although we have presented the uncertainty of each link, users are not clear how the model finds these links. We hypothesize that using visualizations to explain the causal link can significantly improve users' confidence about the causal detection result and thereby facilitate the causal analysis. We have considered two different solutions for addressing this issue. One is to visually present specific cases in the raw data for supporting the detection result and the other is to visualize the causal detection process.
The analysts also suggested that showing the ``deleted correlations'' is potentially helpful for the causal understanding. Explaining why certain relations are considered as correlation but not causation may help users understand the detection process which provides guidance for the link validation.

\textbf{Applications of Causal Analysis.} The second implication is to apply causal analysis to the user segmentation and the comparative analysis of user groups. User segmentation is to segment users into explainable groups according to their characteristics, such as ages, genders, and regions. Applying causal analysis to each group of users can significantly improve the explainability as analysts can clearly state the group difference by comparing the causal links. However, this application is blocked by the causal discovery algorithms. User segmentation is usually an interactive process that cannot be supported by existing causal discovery due to the high time complexity. Parallelizing the causal discovery is a possible solution for this issue.

\textbf{Pitfalls of Causal Detection.} Although many statistical machine learning models have been developed to enable automatic detection of causal relations, it is still difficult to guarantee that every detected causal relation is real and trustworthy. Here, we reflect on our studies and discuss the critical pitfalls that may lead to incorrect and even harmful causal detection results.

Confounding bias is an important pitfall that could impair the accuracy of causal detection. For example, given two independent variables $X$ and $Y$, if they are causally influenced by a third variable $Z$ (confounder), a spurious association between $X$ and $Y$ will be observed. F-GES can handle certain obvious confounders and remove the corresponding spurious associations from the causal graph, which is one of the reasons why we used this model in our system. However, fully addressing the confounding pitfall still remains a difficult problem, especially when the confounders are not observed in the data.

Data quality is a general issue in statistical analysis and also has an impact on causal detection. For example, Berkson's paradox~\cite{berkson1946limitations} (i.e., two positively related or even unrelated dimensions being observed as negatively related) is a phenomenon caused by data selection biases and can lead to incorrect causal links. Recent experiments~\cite{shen2020challenges} suggest that providing more data dimensions and more prior knowledge of the relationships between dimensions can reduce incorrect causal links. However, in many real-world scenarios, adding more data dimensions leads to a smaller sample size, which will decrease the statistical power of causal detection and lower the number of detected causal links.

Moreover, causal detection becomes more complex when temporal dimensions are included. Many new issues are introduced that can lead to incorrect causal links. For example, the sampling rate of the data may not match the changing rate of the temporal dimension, the causal relations may evolve and change dramatically over time, and lagged causal effects may also exist.

Due to the aforementioned pitfalls, the performance of automated causal detection cannot always be guaranteed.  One promising solution is to keep humans in the loop of causal analyses to review results and make trade-offs for mitigating the pitfalls, or conduct controlled experiments to eventually confirm the cause effects.

\textbf{Limitations.}
We identified two limitations in our work. The first limitation is the neglect of temporal variables. Variables in our cases are all static. However, it is common to have temporal variables in domain applications, such as users' online clickstreams.
% To analyze the causality of temporal variables, such as the causal impact of users' different steps of clicking behavior, 
One solution to support temporal variables is to use a Dynamic Bayesian Network (DBN)~\cite{dbn}, where nodes could contain temporal information. We can therefore adapt our approach to DBN by transforming the temporal nodes of DBN to static nodes. However, there are many issues remained to be addressed. For example, the causal links between temporal variables and static variables should be distinguished by different encodings, the uncertainty information of the temporal variables needs to be extracted, and the design requirements for performing what-if analysis on temporal variables need to be gathered. We plan to address these challenges and support temporal variables in our follow-up research.

The second limitation is about the integration of users' domain knowledge. In this study, the causal network is automatically detected. Although this is useful for the causal analysis of high dimensional data, users are still willing to have a solution for interactively adding their self-defined causal links, re-computing the causal graph, and further conducting what-if analysis with the new graph. This can fully utilize users' domain knowledge and create an efficient analytic loop, i.e., users obtain insights from the causal graph and in turn guide the graph detection by feeding their insights. With the development of causal detection models, we can address this limitation in the future.

\section{Conclusion}

In this work, we have identified the key challenges and user needs for conducting exploratory causal analysis through interviews with 5 practitioners. We have designed and implemented a visual analytics system that features a scalable causal graph layout for causal exploration and a set of user interactions for what-if analysis. We have conducted two case studies with experts from education and marketing domains to evaluate the usability and effectiveness of the system. The case studies suggest that the system is easy to learn and use, the causal graph layout is readable even when showing a large set of causal relations, and the what-if analysis is useful for making action plans and estimating the impact. In the future, we will extend our approach to support temporal variables in causal analysis. We will also adapt our system workflow to incorporate human knowledge into the causal discovery process. Finally, we plan to formally evaluate our uncertainty aware visualization of causal relations through controlled user studies.

\acknowledgments{The authors wish to thank all the reviewers, study participants, and domain experts for their thoughtful feedback. The work was supported by National Key R\&D Program of China (2018YFB1004300 ), NSFC (61761136020), NSFC-Zhejiang Joint Fund for the Integration of Industrialization and Informatization (U1609217), Zhejiang Provincial Natural Science Foundation (LR18F020001) and the 100 Talents Program of Zhejiang University.}

\end{spacing}

\bibliographystyle{abbrv-doi}

\bibliography{main}
\end{document}